\documentclass{ws-procs975x65}

\newcommand{\hmpc}{$h^{-1}${\rm Mpc}}

\begin{document}

\title{\uppercase{The Phase-Space Structure of Cold Dark Matter in the Universe}}

\author{\uppercase{Sergei Shandarin}}

\address{Department of Physics and Astronomy, University of Kansas\\
Lawrence, Kansas 66045, USA\\
E-mail: sergei@ku.edu\\
}

\begin{abstract}
A novel method allowing to compute  density, velocity  and other fields in cosmological N--body simulations 
with unprecedentedly  high spatial  resolution is described. 
It is based on the tessellation of the three-dimensional manifold representing cold dark matter in 
six-dimensional phase space. 
The density, velocity and other fields are computed by  projecting the tessellation on 
configuration space. The application of this technique to cosmological N--body simulations
in $\Lambda$CDM cosmology reveals a far more elaborate cosmic web  then dot plots or
self--adaptive SPH. 
In addition, this method allows to uniquely define physical voids and identify and study the caustic surfaces directly.
\end{abstract}

\keywords{Cosmology: large-scale structure of universe;  numerical simulations}

\bodymatter

\bigskip

Cosmological N-body simulations  currently represent the major  tool for the theoretical studies of the
formation and evolution of the cosmic web. They provide the most accurate statistical information 
on the distribution of mass and galaxies to be compared with observations. Most of the simulations deal with
the growth of the structure only in collisionless dark matter  (hereafter DM) since it dominates
the mass  being able to cluster gravitationally. 
Due to a huge number of dark matter particles the most accurate physical
model of DM is collisionless fluid. 

A new method\cite{shh12} (see also \cite{ahk12})  considers the growth  of cold dark matter web
 as the evolution of a three--dimensional manifold
(the phase--space sheet, hereafter PSS) in six--dimensional phase space.
The degeneracy of the PSS is a result  of extremely  low thermal velocity dispersion in cold DM.
The PSS is approximated by a tessellation with the vertices  represented by the particles of N-body simulation.
The tessellation must be created at the linear stage before shell crossing occurs 
and must remain  intact throughout the rest of the evolution. 
Out of many choices the tetrahedral tessellation  is the most sensible because tetrahedra practically always 
remain convex. 
The exceptions happen only at isolated instances of time when four vertices  become  coplanar 
and the tetrahedron degenerates into a flat polygon. 
Computing the volume of the tetrahedron at later times
using the initial order of vertices results in a negative  value. If the tetrahedron
passes through the coplanar state one more time its volume  become positive again.
These metamorphoses can repeat many times.

The vertices aka particles are treated as the flow tracers while the tetrahedra as the mass tracers.
The density of each tetrahedron can be easily computed assuming  that the mass of each tetrahedron 
is conserved and uniformly distributed within its volume. The tetrahedra tile the Eulerian configuration space 
without overlapping in the linear  regime before shell crossing occurs. 
At later times they start to overlap  resulting in the origin of multi--stream flow regions. 
Various fields (density, velocity and others) can be derived by projecting the PSS on configuration space.
This  techniques uses the full phase--space information  available in the simulations. 
As  a result the derived fields can be computed with unprecedented spatial resolution. 
In addition,  new fields such as multi-stream\cite{s11}
and parity\cite{n12,ns12} fields can also be obtained. 
These  fields provide 
new invaluable information about the dynamics  of the cosmic web which is
supplemental to  the density and velocity fields routinely  used in cosmology.
In particular, the whole volume with one--stream flow can be defined as the physical void 
i.e. the volume devoid of structures that the N--body simulation is capable to resolve.

The signs of the tetrahedra volumes make the parity field in Lagrangian space\cite{n12}. 
This field has a nontrivial topological property: the three--dimensional map of parity can 
be painted by only two different colors. 
The interface between regions with different colors approximates the caustic surfaces.
The number of streams is an odd integer at every generic point of configuration space. 
The number of stream can be even only on caustic surfaces having measure zero. 
The number of streams can is easy to compute at an arbitrary 
point by counting the number of tetrahedra enclosing it. Similarly the density and
other fields can be estimated by summing up the corresponding  quantities over all tetrahedra  enclosing
the chosen point.
 
We demonstrate the method using an example from N-body
simulation of the $\Lambda$CDM cosmological model in a 512\hmpc-sided cubic box\cite{shh12}.  
The number of particles is 512$^3$ and the grid size in the gravitational Poisson
solver is 1024$^3$. The parameters of the $\Lambda$CDM
model are as follows: $h = H_0/(100 \, {\rm km/s}\cdot {\rm Mpc}) = 0.72$,
$\Omega_{tot} = 0.25$, $\Omega_b= 0.043$, $n = 0.97$, $\sigma_8 =
0.8$, the initial redshift $z_{\rm in} = 200$.  

We present a relatively small box cut from a much larger simulation box
in order to reduce the obscuration effects.
Figure 1 illustrates the change in the appearance of the structure with the growth of spatial resolution
of the sampling grid employed for computing the fields.
It is worth stressing that the number of particles and their coordinates are exactly same in all 
panels. The structures in density  (top panels) and multi--stream (bottom panels) fields obviously have
some common features however they are not identical (see also\cite{s11}). 

The left panels show the structure at the mass resolution 
of the N-body simulation. It is remarkable that the multi--stream field clearly demonstrates pancakes
at the lowest spatial resolution while the density field hardly suggests even a hint of the pancake presence.
The explanation of the failure of detecting pancakes in N--body simulations was suggested long time
ago\cite{ks83,sz89}. The pancakes contain a low fraction of mass and at the same time occupy a relatively 
large volume that results in a very low density contrast. 
This was quantitatively confirmed in the recent analysis of
cosmological N--body simulations by applying a new elaborate technique called 
Multiscale Morphology Filter\cite{awj10}. The mean density contrast of pancakes aka walls 
the $\Lambda$CDM cosmological model is about 1.1
while that of filaments and clusters is 4.5 and 73 respectively.

The panels in the middle and on the right demonstrate the structure rendered with four and sixteen times 
higher resolution respectively.  They clearly show all essential components of the cosmic web: clusters, filaments, 
pancakes, and voids. 
\begin{figure}
\begin{center}
  \parbox{0.9\textwidth}{\epsfig{figure=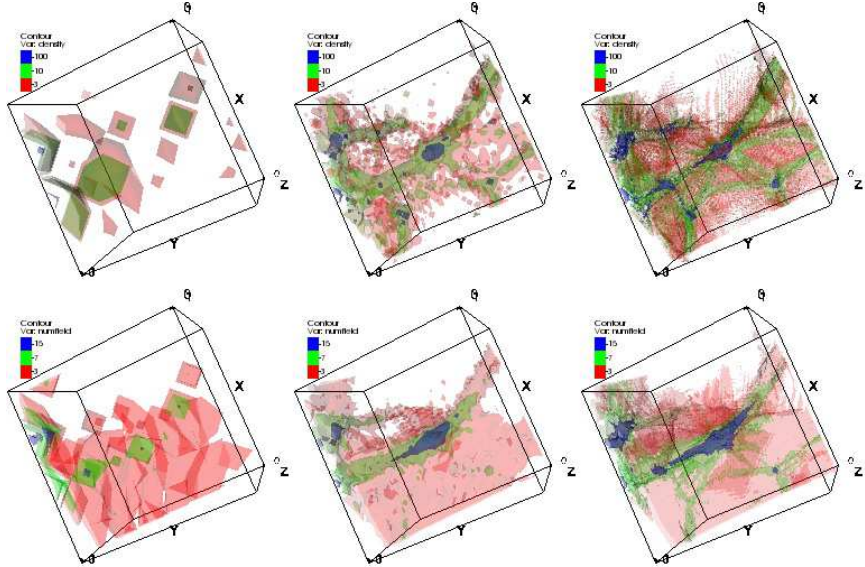,width=0.9\textwidth}}
  \hspace*{0.01\textwidth} 
  \caption{A small 8\hmpc-sided  box. Top row: Density field; Bottom row: Multi-stream field showing the number of streams at every point.
  Three panels in each row correspond to rendering with increasing spatial resolutions: 
  left -- 1 Mpc/h, center -- 0.25 Mpc/h, right -- 0.0625 Mpc/h. }
  \label{fig1}
\end{center}
\end{figure} 

Summarizing we stress the major points. The phase--space sheet contains significantly more dynamical information than coordinates and
velocities of particles if particles are treated as independent items. The connectivity of the tessellation
allows to recover underlying structure much more accurately and with much greater spatial resolution than dot plots or even self--adaptive  SPH\cite{ahk12}. 
Instead of using the particle discretization of the PSS the suggested method is based
on a piecewise-linear approximation.
The new
technique allows to reveal, visualize and quantify the complexity of the cosmic web at a much deeper and
more profound level.

\section*{Acknowledgment}
{\small The author is pleased to acknowledge the support from the New Frontiers in Astronomy and Cosmology program at John Templeton Foundation.}
 



\bibliographystyle{ws-procs975x65}

\end{document}